\documentclass[12pt,a4paper]{article}

\def\blankcite{\@ifnextchar 
[{\@tempswatrue\@blankcitex}{\@tempswafalse\@blankcitex[]}}
\def\@blankcitex[#1]#2{%
  \let\@citea\@empty
  \@blankcite{\@for\@citeb:=#2\do
    {\@citea\def\@citea{,\penalty\@m\ }%
     \edef\@citeb{\expandafter\@iden\@citeb}%
     \if@filesw\immediate\write\@auxout{\string\citation{\@citeb}}\fi
     \@ifundefined{b@\@citeb}{{\reset@font\bfseries ?}%
       \G@refundefinedtrue\@latex@warning
       {Citation `\@citeb' on page \thepage \space undefined}}%
     {\hbox{\csname b@\@citeb\endcsname}}}}{#1}}
\def\@blankcite#1#2{{#1\if@tempswa , #2\fi}}

\newfont{\mybb}{msbm10 scaled 1200}

\newcommand{\ind}{\hspace{.5cm}}

\newcommand{\be}{\begin{equation}}
\newcommand{\ee}{\end{equation}}
\newcommand{\bea}{\begin{eqnarray}}
\newcommand{\eea}{\end{eqnarray}} 
\newcommand{\nn}{\nonumber}

\newcommand{\m}{{\scriptscriptstyle -}}
\newcommand{\p}{{\scriptscriptstyle +}}

\newcommand{\ppint}{\int\!\!\!\!\!\!-}

\newcommand{\bra}{\langle}
\newcommand{\ket}{\rangle}

\newcommand{\cond}{\bra 0 | \bar \psi \psi | 0 \ket}

\newcommand{\pad}[2]{\frac{\partial #1}{\partial #2}}

\newcommand{\comm}[2]{\Big[#1 , #2 \Big] }
\newcommand{\vc}[1]{\mbox{\bf #1}}

\newcommand{\CH}{{\cal H}}

\renewcommand{\thefootnote}{\alph{footnote}}

\begin{document}

\thispagestyle{empty}

\hfill \parbox{4cm}{TPR-96-07 \\
                    hep-th/9608043}
\vspace{1cm}

\begin{center}

{\huge Fermion Condensates  and the Trivial  Vacuum \\[5pt]
of Light-Cone Quantum Field Theory}

\vspace{1.5cm}

{\large 
T.~Heinzl\footnote{e-mail: thomas.heinzl@physik.uni-regensburg.de} \\ 
    
Universit\"at Regensburg \\   
Institut   f\"ur  Theoretische   Physik \\  
93053 Regensburg, FRG }

\end{center}

\vspace{1.5cm}

\begin{abstract}
We  discuss  the  definition  of  condensates  within  light-cone
quantum field theory.  As the vacuum state in this formulation is
trivial,  we  suggest  to abstract  vacuum  properties  from  the
particle spectrum.  The latter can in principle be calculated  by
solving the eigenvalue problem of the light-cone Hamiltonian.  We
focus  on fermionic  condensates  which are order  parameters  of
chiral  symmetry  breaking.   As a paradigm  identity  we use the
Gell-Mann-Oakes-Renner  relation between the quark condensate and
the  observable  pion  mass.   We examine  the analogues  of this
relation  in the `t~Hooft and Schwinger  model, respectively.   A
brief  discussion  of  the  Nambu-Jona-Lasinio  model  is  added.
\end{abstract}

\vfill
\newpage

\renewcommand{\thefootnote}{\arabic{footnote}}
\setcounter{footnote}{0}
\setcounter{page}{2}

{\large \bf 1.} In two recent publications \cite{CZ95, Bur96} the
vacuum structure  of the `t~Hooft model \cite{tHo74},  QCD in 1+1
dimensions  in the limit of large $N_C$, was analysed in terms of
light-cone  (LC) wave functions.  These had already been obtained
in  `t~Hooft's   original   work  \cite{tHo74}   by  solving  the
associated   Bethe-Salpeter   equation  which  is  equivalent  to
diagonalising   the  LC  Hamiltonian.   A  remarkable  result  of
\cite{CZ95,  Bur96}  is the fact  that  the quark  condensate,  a
quantity   indicating   a  non-trivial   vacuum  structure,   can
efficiently  be calculated  in the framework  of LC quantum field
theory  which  is generally  believed  to have  a trivial  vacuum
\cite{LKS70, Rob96}.

\ind
The purpose  of this note  is to further  clarify  this  apparent
contradiction  and put the definition  of condensates  within  LC
field theory in a broader perspective.

\vspace{.5cm}

{\large  \bf  2.}  Assume  that  we  have  a  symmetry  which  is
explicitly   broken  so  that  the  associated   current  is  not
conserved,

\be
\partial_\mu j^\mu (x) = A (x) \; .
\ee

With the help of the corresponding  Ward identity \cite{CJ71} one
can derive the following formula for an arbitrary operator $B$,

\be
\bra 0 | \delta  B / \delta  \alpha | 0 \ket = -i \int d^{\,4}  x
\bra 0 | T A(x) B (0) | 0 \ket = - \sum_n \frac{\bra 0 | A(0) |
n \ket \bra n | B(0) | 0 \ket}{m_n^2} \; .
\label{MASTER}
\ee

Here, $\delta B / \delta \alpha$  denotes the change of $B$ under
the symmetry transformation,  and in the last step a complete set
of states $ |n \ket$, each of mass $m_n$, has been inserted.  For
the case of chiral symmetry, choosing

\be 
A = B = 2 m \bar \psi i \gamma_5 \psi \; ,
\ee

one finds for a single quark flavour of mass $m$

\be
\bra  0 |\bar  \psi \psi | 0 \ket = -m \sum_n  \frac{|\bra  0 |
\bar \psi i \gamma_5 \psi | n \ket|^2}{m_n^2} \; .
\label{COND1}
\ee
 
In Ref.s  \cite{CZ95,  Bur96},  this  expression  was  used  as a
definition  for the quark condensate in the `t~Hooft model.  From
the above  derivation,  however,  it is clear  that (\ref{COND1})
holds  quite  generally.   Using the PCAC relation  for the axial
vector current, $j_5^\mu$,

\be
\partial_\mu j_5^\mu (x) = f_\pi m_\pi^2 \, \pi (x) \; ,
\label{PCAC}
\ee

where  $\pi (x)$ is an interpolating  pion field and $f_\pi$  the
pion decay constant,  (\ref{MASTER})  involves the pion two-point
function and is easily evaluated with the result

\be
f_\pi^2  m_\pi^2  = - 4 m \bra 0 | \bar \psi \psi | 0 \ket \; .
\label{GOR1}
\ee

This  is  the  famous   Gell-Mann-Oakes-Renner   (GOR)   relation
\cite{GOR68} (to lowest order in the quark mass $m$ \cite{GL84}),
which relates  the QCD parameters,  $m$, the current  quark mass,
and  the  fermion  condensate  to  the  observables  $f_\pi$  and
$m_\pi$.  We have written everything  in terms of bare quantities
since the right-hand  side does not change  under renormalisation
\cite{PT84}.   Thus, $m_\pi$ in (\ref{GOR1}) is the physical pion
mass.

\ind
The   GOR  relation   can  of  course   also   be  derived   from
(\ref{COND1}) if one replaces

\be
\bar \psi (x) i \gamma_5 \psi (x) = \frac{1}{2m} f_\pi m_\pi^2 \,
\pi (x) \; , 
\ee

and assumes that, for small $m_\pi$, the sum is saturated  by the
pion.

\ind
In any case, we want to stress the fact that in (\ref{COND1}) and
(\ref{GOR1})   above  a  vacuum  quantity,  the  condensate,   is
expressed  in terms  of the  particle  spectrum.   So,  once  the
spectrum  is known, after, say, diagonalising  the LC Hamiltonian
by one of the various  methods on the market \cite{BP85},  we can
translate back properties  of the spectrum into properties of the
vacuum.   In view of that, we suggest to deemphasize  the role of
the vacuum,  which  is natural  to the extent  that  most  of its
properties are not directly observable. This is particularly true
within the LC framework, where the vacuum state seems to decouple
completely from the particle states.  Similar ideas have been put
forward  long ago, in the context of chiral symmetry  in the (LC)
parton  model,  by  Susskind  et  al.~\cite{CNS71}:    ``In  this
framework  the spontaneous  symmetry breakdown must be attributed
to the properties  of the hadron's  wavefunction  and  not to the
vacuum'' \cite{CS74}. A related point of view has also been taken
more recently in \cite{LLT91}.

\vspace{.5cm}

{\large \bf 3.}
Before  we pursue  the program  just  outlined  we would  like to
remark  that  the  `master  equation'  (\ref{MASTER})  cannot  be
derived by strictly  sticking  to the LC framework.   To this end
note that the first  term in (\ref{MASTER})  can be written  with
the help of the charge

\be
Q(x^0) = \int d^{\,3} x \, j^0 (x^0 , \vc{x} ) \; ,
\ee

the generator of the symmetry, as

\be
\bra  0  |  \delta  B  /  \delta  \alpha  | 0 \ket  = -i  \bra  0
|\comm{B}{Q} | 0 \ket \; ,
\label{CHARGECOMM}
\ee

where the commutator  is evaluated at equal time $x^0 = 0$.  This
expression  cannot be directly translated into the LC language by
replacing   the   ordinary   charge   $Q(x^0)$   by  the   LC  or
``light-like'' charge,

\be
Q(x^\p) = \int dx^\m d^2x_\perp \, j^\p (x^\p, x^\m, \vc{x}_\perp
) \; ,
\ee

and  evaluating   the  commutator   at  equal   {\em  light-cone}
time\footnote{our   LC  conventions   (with   $a$   an  arbitrary
four-vector) are:  $a^\pm = (a^0 \pm a^3)/\sqrt{2}$, $(a^1 , a^2)
= \vc{a}_\perp$.},  $x^\p$.   The reason  for this  is a peculiar
property of LC charges:  they annihilate the vacuum, irrespective
of whether they generate  a symmetry  or not \cite{LKS70,  JS68},
which  is in accordance  with  the  triviality  of the LC vacuum.
Thus, the right-hand  side of (\ref{CHARGECOMM}),  evaluated on a
null-plane,  is always  zero.   Hence,  that part of the operator
$\delta  B  /  \delta  \alpha$  having  a  non-vanishing   vacuum
expectation  value (VEV) cannot  be obtained  by an infinitesimal
transformation  generated  by  the  light-like  charge  $Q$.  For
example, in the LC sigma model, the relation

\be
\comm{\pi}{Q_5} = -i \sigma \; 
\label{SIGMA-COMM}
\ee

does only hold for those modes of the field operators,  $\pi$ and
$\sigma$,  having  non-vanishing   LC  three-momenta,   $(p^\p  ,
\vc{p}_\perp ) \ne 0$ \cite{Hei92}.   These non-zero modes do not
have a VEV.  Thus, the VEV of (\ref{SIGMA-COMM}) vanishes on both
sides, as it should.

\ind
The moral  is that we have to {\em  assume}  the validity  of the
identity  (\ref{MASTER})  for any possible choice of quantisation
hypersurface,  in particular  for  one  tangent  to the  LC, {\it
i.e.}~a null plane.  This means in particular,  that we can use a
complete set of eigenstates  of the {\em light-cone}  Hamiltonian
in the last term  of (\ref{MASTER}),  as was done  in \cite{CZ95,
Bur96}.

\vspace{.5cm}

{\large \bf 4.} The `t~Hooft  model is the simplest  theory where
use can be made  of LC wave  functions.   This is due to the fact
that, in the limit of large $N_C$,  the coupling  to higher  Fock
states is suppressed by negative powers of $N_C$.  Therefore, the
LC eigenvalue  problem closes in the quark-anti-quark  sector and
the sum in (\ref{COND1})  is saturated  by one-meson states.  For
the matrix elements in (\ref{COND1}) one finds \cite{CZ95, Bur96,
CCG76, Zhi86}

\be
F_n  \equiv  \bra  0  | \bar  \psi  i \gamma_5  \psi  | n \ket  =
\sqrt{\frac{N_C}{\pi}}  \frac{m}{2}  \int_0^1  dx \, \frac{\phi_n
(x)}{x(1-x)} \; . 
\label{THO1} 
\ee

The $\phi_n(x)$  are the LC wave functions  of the mesonic  bound
states and are obtained as solutions of `t~Hooft's equation,

\be
m_n^2 \phi_n  (x) = \frac{m^2}{x  (1-x)}  \phi_n  (x) + \mu_{0}^2
\ppint dy \frac{\phi_n(x) - \phi_n (y)}{(x-y)^2} \; .
\label{THO-EQ1}
\ee

The variables $x,y$ represent the LC momentum fraction carried by
one of the quarks  in the meson,  the integral  on the right-hand
side  is  performed  with  a principal  value  prescription,  and
$\mu_{0}^2  = g^2  N_C /2\pi$  is the  basic  mass  scale  of the
theory.  Integrating  (\ref{THO-EQ1})  over  $x$  one  finds  the
important relation \cite{CCG76}

\be
m_n^2  \int_0^1  dx  \,  \phi_n  (x)  =  m^2  \int_0^1   dx  \,
\frac{\phi_n (x)}{x(1-x)} \; . 
\label{INT-INT}
\ee

The left-hand  side is basically  the wave function at the origin
($x^\m  = 0$),  which  is  expressed  in  terms  of  an  integral
dominated  by the  infrared  tails,  $x \to$  0 or 1.  

\ind
In the  chiral  limit,  $m \to 0$, the  sum  in (\ref{COND1})  is
saturated  by the meson of lowest  mass,  which  we will call the
`pion'. The condensate is thus given by

\be
\bra 0 | \bar \psi \psi | 0 \ket = - \frac{m F_\pi^2}{m_\pi^2} \;
,
\label{COND2}
\ee

and one has to calculate  $F_\pi$ from (\ref{THO1})  and $m_\pi$.
This has been done in \cite{Zhi86}  using `t~Hooft's  ansatz  for
the wave function,

\be
\phi_\pi (x) \simeq x^\beta (1-x)^\beta \; ,
\label{THO-ANS}
\ee

where   $\beta$   is   determined   from   `t~Hooft's    equation
(\ref{THO-EQ1}), yielding

\be
\beta \, \frac{\pi}{\sqrt{3}} = \frac{m}{\mu_{0}} \; .
\label{BETA}
\ee

We have explicitly  displayed  the factor  $\pi/\sqrt{3}$,  which
will play a peculiar role later.  Inserting (\ref{THO-ANS})  into
(\ref{INT-INT}), one obtains for the `pion' mass

\be
m_\pi^2 = \frac{2m^2}{\beta} = 2 \, \frac{\pi}{\sqrt{3}} \, m \mu_{0}
\; ,
\label{PI-MASS1}
\ee

and for $F_\pi$,

\be
F_\pi = \sqrt{N_C / \pi} \, \frac{\pi}{\sqrt{3}} \, \mu_{0} \; .
\label{FPI}
\ee

Inserting  (\ref{PI-MASS1})  and (\ref{FPI})  into (\ref{COND2}),
one easily finds the condensate \cite{CZ95, Bur96, LLT91, Zhi86}

\be
\bra 0 | \bar \psi \psi | 0 \ket = - \frac{N_C}{\sqrt{12}} \, \mu_{0}
= - \frac{N_C}{2\pi} \, \frac{\pi}{\sqrt{3}} \, \mu_{0} \; .
\label{COND3}
\ee

As  was  to  be  expected,  all  three  dimensionful  quantities,
$m_\pi$,  $F_\pi$  and $\bra  0 | \bar  \psi  \psi | 0 \ket$, are
expressed  in terms  of the basic scale  $\mu_{0}$.   It is, however,
interesting  to write  these  in terms  of the parameter  $\beta$
characterising the LC wave functions. One obtains

\bea
m_\pi^2 &=& 2 m \, \frac{m}{\beta} \; , \label{PI-MASS2} \\
F_\pi &=& \sqrt{N_C / \pi} \, \frac{m}{\beta} \; , \\
\cond &=& - \frac{N_C}{2\pi} \, \frac{m}{\beta} \label{COND4}\; .
\eea

Thus,  all `low energy  parameters'  are expressible  in terms of
$\beta$ and the quark mass $m$ which are equivalent to $\mu_{0}$,
as is evident from (\ref{BETA}).

\ind
One last way of expressing  the low energy parameters is in terms
of dimensionless fractions, {\it i.e.}~by measuring all masses in
units of the scale $\mu_{0}$,

\bea
\frac{m_\pi^2}{\mu_{0}^2}   &=&   2   \,   \frac{m}{\mu_{0}}   \,
\frac{\pi}{\sqrt{3}}\; , \\
\frac{F_\pi}{\mu_{0}}     &=&     \sqrt{N_C     /     \pi}     \,
\frac{\pi}{\sqrt{3}} \; , \\
\frac{\cond}{\mu_{0}}      &=&     -     \frac{N_C}{2\pi}      \,
\frac{\pi}{\sqrt{3}} \; .
\eea

The  above  way of writing  everything  once  more  exhibits  the
mysterious factor $\pi/\sqrt{3}$. We will come back to this issue
later.

\ind
Finally,  to make contact with the GOR relation (\ref{GOR1}),  we
insert (\ref{COND4})  into (\ref{PI-MASS2})  which yields the GOR
relation in the `t~Hooft model,

\be
\frac{N_C}{\pi} m_\pi^2 = - 4 m \cond  \; .
\label{THO-GOR}
\ee

This leads to yet another definition of the condensate, namely in
terms of the `pion' mass,

\be
\cond = - \frac{N_C}{4\pi}  \left.  \pad{}{m}  m_\pi^2 (m) \right
\vert_{m= 0} \; .
\label{COND5}
\ee

In contrast  to the GOR relation  (\ref{THO-GOR})  which holds to
first  order  in the quark  mass  $m$,  (\ref{COND5})  is an {\em
exact} statement.

\vspace{.5cm}

{\large  \bf  5.}  The  massive  Schwinger   model  (QED  in  1+1
dimensions  \cite{Sch62,  CJS75, Col76})  has first been analysed
along  the  lines  of  `t~Hooft  by Bergknoff  \cite{Ber77},  who
calculated  the bound  state spectrum  using  LC methods.   These
results  have  been  refined  and  extended  recently  by several
authors  \cite{MP93,  HOT95}\footnote{There  have  also been many
attempts  to obtain  the exact (operator)  solution  for the {\em
massless}  Schwinger  model within  LC quantisation.   Due to the
singular  nature  of  massless  fields  in  1+1  dimensions,   in
particular  on the  LC, the necessary  efforts  are  considerably
larger   than  within   ordinary   quantisation   on  $x^0  =  0$
\cite{McC91}.}.  In  the  massive  Schwinger  model,  one  cannot
straightforwardly  make use of the relations  (\ref{MASTER})  and
(\ref{COND1})  in order to define the condensate.  This is due to
the fact that the axial current is anomalous \cite{Joh63, Jac84},

\be
\partial_\mu   j_5^\mu   =  -  \frac{e}{2\pi}   \epsilon_{\mu\nu}
F^{\mu\nu} + 2m \, \bar \psi i \gamma_5 \psi \; ,
\label{ANOMALY}
\ee

so that it is {\em not} conserved  in the chiral limit $m \to 0$.
The condensate  can, however, be obtained exactly by bosonisation
of the massless Schwinger model \cite{BKS76, Smi92},

\be
\cond = - \frac{1}{2\pi} \, e^\gamma \, \mu_{0} \; ,
\label{COND6}
\ee

where $\gamma$ = 0.577...  denotes Euler's constant. On the other
hand, the `pion' mass has recently  been calculated  up to second
order  in the fermion  mass $m$ \cite{Ada95},  from which one can
derive   an  expression   for   the   condensate   analogous   to
(\ref{COND5}). We only need the first order result \cite{BKS76},

\bea
m_\pi^2 (m) &=& \mu_{0}^2 - 4\pi m \cond + O(m^2) \nn \\ 
&=& \mu_0^2 + 2 e^\gamma m \mu_0 + O(m^2) \; ,
\label{SM-MPI1}
\eea

where  $\mu_{0}  = e/\sqrt{\pi}$  is the basic mass scale  of the
Schwinger model\footnote{In  (\ref{SM-MPI1}) and the following we
set  the  $\theta$-angle   \cite{CJS75,  Col76}  equal  to  zero.
Including  a non-vanishing  vacuum angle turns out to be a rather
non-trivial   task.   A  first  attempt  has  appeared   recently
\cite{HOT96}.}.   From  (\ref{SM-MPI1})  it is evident  that  the
Schwinger model analogue of (\ref{COND5}) holds, namely

\be
\cond = - \frac{1}{4\pi}  \left.  \pad{}{m}  m_\pi^2 (m) \right
\vert_{m= 0} \; .
\label{COND7}
\ee

All one has to do to obtain (\ref{COND7}) from (\ref{COND5}),  is
to  replace   $N_C$  by  one.   This  suggests   that  expression
(\ref{SM-MPI1}) for the `pion' mass squared can be derived in the
same way as for the `t~Hooft model. In some sense this is not too
surprising,  since the bound state equations of both the `t~Hooft
and the Schwinger  model (in the two-particle  or valence sector)
can be written in a unified way \cite{BHP90},

\be
m_n^2  \phi_n  (x)  = \frac{m^2}{x  (1-x)}  \phi_n  (x)  + \alpha
\mu_{0}^2  \int_0^1  dx  \,  \phi_n  (x)  + \mu_{0}^2  \ppint  dy
\frac{\phi_n(x) - \phi_n (y)}{(x-y)^2} \; .
\label{THO-EQ2} 
\ee

In the `t~Hooft  model, $\alpha = 0$, and in the Schwinger  model
$\alpha  = 1$.   The scale  parameters  $\mu_{0}$  are  given  by
$\mu_{0}^2  = g^2  N_C  / 2\pi$  and  $ \mu_{0}^2  = e^2  / \pi$,
respectively.   By performing  exactly  the same steps as for the
`t~Hooft model one finds the `pion' mass squared \cite{Ber77}

\be
m_\pi^2  = \alpha \mu_{0}^2  + 2 \, \frac{\pi}{\sqrt{3}}  \, m \,
\mu_{0} \; .
\label{SM-MPI2}
\ee

In this expression,  the first term on the right-hand side is due
to the anomaly.   The  parameter  $\alpha$  thus  `measures'  the
strength  of the anomaly,  which in the `t~Hooft  model is absent
($\alpha$ = 0). Comparing (\ref{SM-MPI1}) and (\ref{SM-MPI2}) one
notes,  however,  a difference:   in the second  expression,  the
factor $e^\gamma$  is replaced by the ubiquitous  $\pi/\sqrt{3}$.
Numerically, one has

\be
\pi/ \sqrt{3} = 1.814 \; , \quad e^\gamma = 1.781 \; ,
\label{DISCREP}
\ee

so that the difference is about 2\% \cite{Ber77}.   There are two
possible  sources  for this discrepancy.   First,  there might be
contributions from higher Fock sectors. These, however, vanish in
the large-$N_c$  limit (`t~Hooft  model) as well as in the chiral
limit (Schwinger  model).   In both these limits,  the associated
`pion' is exactly two-particle.  The second approximation  is the
use of `t~Hooft's  ansatz  (\ref{THO-ANS})  for the wavefunctions
which yields a good description  for the endpoint  behaviour,  $x
\to 0$ or $1$, but not for intermediate values of $x$.  While the
error should be small as the condensate {\em is} dominated by the
endpoint  behaviour,  we nonetheless  believe that neglecting  the
non-asymptotic  regions  in $x$ is the main  reason  for  the 2\%
discrepancy. In principle, this can be checked by using numerical
methods  like those of Harada  {\em et al.}~\cite{HOT95}.   These
authors  went much beyond `t~Hooft's  ansatz  (using an elaborate
wave function basis) and included up to six-particle  states.  In
practice,  however,  it  turns  out  to  be rather  difficult  to
precisely  determine  the small-$m$ behaviour of $m_\pi^2$ as the
numerical   data  do  not  converge  very  well  in  this  region
\cite{Har96}.

\ind
Using    (\ref{COND7})    to   obtain    the   condensate    from
(\ref{SM-MPI2}), one finds

\be
\cond = - \frac{1}{2\pi} \, \frac{\pi}{\sqrt{3}} \, \mu_{0} \; ,
\label{COND8}
\ee

which  is  equivalent  to  (\ref{COND3})  after  the  appropriate
replacements.    Interestingly,   the  condensate  is  completely
independent   of  the  anomaly.   Switching   off the  latter  by
(artificially)  putting  $\alpha$  equal  to zero  also  for  the
Schwinger model, would not change the value of $\cond$.

\ind 
Finally,  we  would  like  to remark,  that, if the  solution  of
the Schwinger  model  bound state  equation  yields  a condensate
which is off by 2\%, there is no reason  to believe  that exactly
the same procedure  yields a correct value for the condensate  in
the  `t~Hooft  model.   Thus  we  conjecture,   that  the  factor
$\pi/\sqrt{3}$ should be replaced by $e^\gamma$ {\em everywhere}.
In particular, the `t~Hooft model condensate (\ref{COND3})  would
then become

\be
\cond = - \frac{N_C}{2\pi} \, e^\gamma \, \mu_{0} \; .
\label{COND9}
\ee

\vspace{.5cm}

{\large  \bf  6.}  One remaining  question  is whether  the above
discrepancy  can  be regarded  as physical.   To answer  this  we
concentrate  on the Schwinger model ($\alpha$=1)  in what follows
and write for the condensate (\ref{COND8}) more generally

\be
\cond = - c \mu_{0} \; ,
\label{COND10}
\ee

and for the `pion' mass (\ref{SM-MPI2}) in units of $\mu_0$,

\be
\frac{m_\pi}{\mu_0} = 1 + 2\pi c \frac{m}{\mu_{0}} + O(m/\mu_0)^2
\; .
\label{SM-MPI3}
\ee

The  question  then  can  be  reformulated:   is there  a unique,
physical value for $c$?  In a recent publication \cite{SW96}, for
example, the authors claim that the constant $c$ is dependent  on
the renormalization scheme used.  They do not, however, give this
dependence explicitly.

\ind
A convenient way to analyse this issue is the following.   We use
the bosonised (sine-Gordon) Hamiltonian (density) \cite{Col75}

\bea
\CH  (m,  \mu_0)  &=&  N_{\mu_0}   \left[  \frac{1}{2}   
\left(\pad{\phi}{t}\right)^2+
\frac{1}{2}\left(\pad{\phi}{x}\right)^2  + \frac{1}{2}  \mu_{0}^2
\phi^2 - c m \mu_{0} \cos (\sqrt{4\pi} \phi) \right] \nn \\
&\equiv& N_{\mu_0} \left[ \CH_0 + \frac{1}{2}  \mu_{0}^2
\phi^2 - c m \mu_{0} \cos (\sqrt{4\pi} \phi) \right] \; ,
\label{SM-HAM1} 
\eea

where $N_{\mu_0}$  denotes  normal-ordering  with respect  to the
scale  $\mu_{0}$.   The  energy  density  in the  vacuum  of mass
$\mu_{0}$ is then

\be
\mathcal{E} (m , \mu_{0}) = \bra 0 , \mu_{0} | \CH (m, \mu_0 )| 0
, \mu_{0} \ket = - c m \mu_{0} \; . 
\ee

This expression can be used to give yet another definition of the
condensate  \cite{CHK82}.   Regarding  the fermion  mass $m$ as a
parameter, the Feynman-Hellmann theorem leads to

\be 
\cond_{\mu_0}  =  \pad{}{m}  \mathcal{E}  (m  ,  \mu_{0}  ) = - c
\mu_{0} \; ,
\label{COND11}
\ee

which is the same as (\ref{COND10}).   What happens  when the the
scale is changed, say, from $\mu_{0}$ to $\mu$? The answer can be
obtained   from   Coleman's   work   on  the  sine-Gordon   model
\cite{Col75}, where one finds

\bea
\CH (m, \mu_0 ) &=& N_{\mu} \left[\CH_0  + \frac{1}{2}  \mu_{0}^2
\phi^2  - c m \mu \cos  (\sqrt{4\pi}  \phi)  \right]  \nn  \\ 
&-& \frac{1}{8\pi}      \left(\mu_{0}^2      +     \mu_{0}^2      \ln
\frac{\mu^2}{\mu_{0}^2} - \mu^2 \right) \; . 
\label{SM-HAM2} 
\eea

The vacuum energy density becomes

\be
\mathcal{E}  (m , \mu) = \bra 0 , \mu | \CH (m, \mu_0  )| 0 , \mu
\ket  =  -  \frac{1}{8\pi}   \left(\mu_{0}^2   +  \mu_{0}^2   \ln
\frac{\mu^2}{\mu_{0}^2} - \mu^2 \right) - c m \mu \; .
\label{VACDENS2} 
\ee

yielding the condensate at scale $\mu$,

\be 
\cond_{\mu} = \pad{}{m} \mathcal{E} (m , \mu ) = - c \mu \; .
\label{COND12} 
\ee

The scale dependence  of the condensate  is thus very simple  and
can  be expressed  in terms  of the  renormalisation  group  (RG)
equation,

\be
\pad{}{\mu} \cond_{\mu} = - c \; .
\label{RG}
\ee

The RG invariant quantity, which does not get renormalised,  thus
being the analogue of $m\cond$ in (\ref{GOR1}), is therefore

\be
\frac{\cond_{\mu}}{\mu} = - c \; .
\label{COND13}
\ee

The conclusion  is that the constant $c$ defines the physical, RG
invariant  value of the condensate  and by itself  does {\em not}
depend  on the renormalisation  scale $\mu$.   As the condensates
(\ref{COND6})  and (\ref{COND8})  are evaluated at the same scale
$\mu_0$,  we  regard  the  discrepancy  (\ref{DISCREP})  as being
relevant and calling for an explanation.

\vspace{.5cm}

{\large  \bf  7.} It is obvious  that  in higher  dimensions  the
renormalisation program becomes much more involved, in particular
for the bound  state  equations  yielding  the LC wave  functions
\cite{PW93}.   An exception  from  this rule  might  be effective
theories which need not be renormalisable and are thus meaningful
only below some physical cutoff $\Lambda$. A prominent example is
the Nambu-Jona-Lasinio  (NJL) model \cite{NJL61}  describing  the
dynamical  breakdown  of  chiral  symmetry.  It  has  a  chirally
symmetric  four-fermion   interaction,  but  (beyond  a  critical
coupling, $g > g_c$,) the vacuum breaks chiral symmetry resulting
in   a   non-vanishing   fermion   condensate.    In   mean-field
approximation this condensate determines the mass gap

\be
m - m_0 = - 2g \cond_m \; ,
\label{GAP-EQ}
\ee

between  the current  quarks  with mass $m_0$ and the constituent
quarks with mass $m$ (dynamical mass generation). The constituent
mass  is  obtained  self-consistent\-ly  from  the  gap  equation
(\ref{GAP-EQ}), and this is how all the non-perturbative  physics
enters.  The condensate itself is calculated perturbatively, {\it
i.e.~}in a Dirac vacuum for free fermions of mass $m$. Again, the
Feynman-Hellmann  theorem is very helpful.  Integrating  over all
the one-particle energies of the Dirac sea, one finds

\bea
\cond_m    &=&    \pad{}{m}    \mathcal{E}    (m)   =   \pad{}{m}
\int\limits_{-\infty}^0    \frac{dk^\p}{2\pi}    \int   \frac{d^2
k_\perp}{(2\pi)^2}  \frac{m^2  + k_\perp^2}{2k^\p}   \nn \\
&=&   -\frac{m}{8\pi^3}    \int\limits_0^{\infty}
\frac{dk^\p}{k^\p}  \int  d^2  k_\perp  \; , 
\label{COND14} 
\eea

As it stands,  the integral  is of course divergent  and requires
regularisation.   In the most straightforward  manner one chooses
$m^2/\Lambda  \le  k^\p  \le  \Lambda$  and  $|\vc{k}_\perp|  \le
\Lambda$, so that the condensate becomes

\be
\cond_m  =  - \frac{m}{8\pi^2}  \int\limits_{m^2/\Lambda}^\Lambda
\frac{dk^\p}{k^\p}  \int\limits_0^{\Lambda^2}  d(k_\perp^2)  =  -
\frac{m}{8\pi^2} \Lambda^2 \ln \frac{\Lambda^2}{m^2} \; .
\label{COND15}
\ee

Plugging   this  result  into  the  gap  equation  (\ref{GAP-EQ})
(setting for simplicity  $m_0 = 0$ in what follows) one finds for
the dynamical mass squared,

\be
m^2  (g)  =  \Lambda^2  \exp  \left(-\frac{4\pi^2}{g   \Lambda^2}
\right) \; .
\label{DYN-MASS1}
\ee

The  critical  coupling  is determined  by the vanishing  of this
mass,  $m (g_c)  = 0$,  and from  (\ref{DYN-MASS1})  we find  the
surprising result

\be
g_c = 0 \; .
\label{GC1}
\ee

This  result,  however,   is  wrong  since  one  knows  from  the
conventional treatment of the model that the critical coupling is
finite  of the order  $\pi^2/\Lambda^2$,  both for covariant  and
non-covariant  cutoff  \cite{NJL61}.   In  addition,  it is quite
generally  clear  that in the {\em free}  theory  ($g=0$)  chiral
symmetry  is not broken  (for  $m_0  = 0$) and,  therefore,  this
should not happen for arbitrarily  small coupling,  either.   The
remedy  is once  more  to use  an information  from  the ordinary
calculation  of the condensate.   We translate the non-covariant,
but rotationally  invariant,  three-vector  cutoff, $|\vc{k}| \le
\Lambda$, into LC coordinates \cite{DHS89}, which leads to

\be
0 \le k_\perp^2  \le 2\Lambda  k^\p  - m^2 - (k^\p)^2  \; , \quad
\frac{m^2}{2\Lambda} \le k^\p \le 2\Lambda \; .  
\label{CUTOFF}
\ee

Note  that  the  transverse   cutoff  becomes  a  polynomial   in
$k^\p$. The $k_\perp$-integration thus has to be performed first.
For   the   condensate   this   yields   an  analytic   structure
different from (\ref{COND15}),

\be
\cond_m  =  -  \frac{m}{8\pi^2}   \left(  2\Lambda^2  -  m^2  \ln
\frac{\Lambda^2}{m^2} \right) \; ,
\label{COND16}
\ee

where we have neglected subleading terms in the cutoff $\Lambda$.
From (\ref{COND16}),  one infers the correct cutoff dependence of
the critical coupling,

\be
g_c = \frac{2\pi^2}{\Lambda^2} \; .
\label{GC2}
\ee

The   moral   of   this   calculation   is   that   even   in   a
non-renormalisable   theory   like   the   NJL   model,   the  LC
regularisation prescription is a subtle issue.  In order to get a
physically  sensible  result  the  transverse  cutoff  has  to be
$k^\p$-dependent.   Clearly, this dependence  cannot be arbitrary
but  should   be  constrained   from  dimensional   and  symmetry
considerations.  For renormalisable theories, such arguments have
been given by Perry and Wilson \cite{PW93}. In the example above,
it was (ordinary) rotational invariance that solved the problem.

\ind
The   condensate   (\ref{COND16})   was   already   obtained   in
\cite{DHS89}, where, however, a slightly more complicated cut-off
was used. In that work, the condensate was defined covariantly in
terms of the fermion propagator at the origin, $S_F(x=0)$,

\be
\cond_m  =  -i  \mbox{tr}  S_F(0)  = -i  \frac{m}{4\pi^4}  \int
\frac{d^4k}{k^2 - m^2 + i\epsilon}  \; .
\ee

Performing  the  integration  over  $k^\m$  with  the appropriate
cutoff leads to (\ref{COND16}).

\ind
Within the NJL model, an illustrating analogy to magnetic systems
can be made.  Chiral symmetry corresponds to rotational symmetry,
the vacuum energy density to the Gibbs free energy,  and the mass
$m$ to an external magnetic field.  The order parameter measuring
the rotational  symmetry  breaking  is the magnetisation.   It is
obtained  by differentiating  the free energy with respect to the
external   field.    This   is   the   analogue   of   expression
(\ref{COND14}) as derived from the Feynman-Hellmann theorem.

\ind
It would be very interesting  to relate the NJL condensate  to LC
wave  functions  in the same  spirit  as for the  two dimensional
models  above.   To this end one has to solve  the (pseudoscalar)
bound  state equation  not only for the pion mass (as was done in
\cite{DHS89}  in the chiral  limit)  but also  for the associated
eigenfunctions  of the LC Hamiltonian.  Work in this direction is
underway.

\vspace{1cm}

The author is indebted to E.~Werner  for continuous  interest and
support.  He also thanks K.~Harada, N.~Kaiser, A.~Kalloniatis and
B.~van de Sande for enlightening  discussions.  Special thanks go
to C.~Stern  for reading the manuscript,  as well as to K.~Harada
and  T.~Sugihara   for  providing   and  explaining   unpublished
numerical results.

\end{document}